\documentclass[a4paper,12pt]{article}

\title{Some Problems\\ on the Classical N-Body Problem\footnote{This preprint is published and should be cited as:
A.\ Albouy, H.E.\ Cabral, A.A.\ Santos, Some problems on the classical $n$-body problem, 
{\it Celestial Mechanics and Dynamical Astronomy}, {\bf 113}(4) (2012), pp.\ 369--375.
The original publication is available at www.springer.com. The digital object identifier (DOI) is : 10.1007/s10569-012-9431-1} }

\date{}

\author{ Alain Albouy\footnote{Astronomie et Syst\`emes Dynamiques, IMCCE, UMR 8028 du CNRS, 77, av. Denfert-Rochereau, 75014 Paris, France, albouy@imcce.fr}\quad
 Hildeberto E.\ Cabral\footnote{Departamento de Matem\'atica,
Universidade Federal de Pernambuco, av.\ Prof.\ Luiz Freire, s/n,
Recife, Pernambuco, Brazil, hild@dmat.ufpe.br}\quad Alan A.\ Santos\footnote{Departamento de Matem\'atica, Universidade Federal de Sergipe, av.\ Ol\'{\i}mpio Grande, s/n, Itabaiana, Sergipe, Brazil, alan@ufs.br}\\
}

\begin{document}

\maketitle
\begin{abstract}
Our idea is to imitate Smale's list of problems, in a restricted domain of mathematical aspects of Celestial Mechanics. All the problems are on the $n$-body problem, some with different homogeneity of the potential, addressing many aspects such as central configurations, stability of relative equilibrium, singularities, integral manifolds, etc. Following Steve Smale in his list, the criteria for our selection are: (1) Simple statement. Also preferably mathematically precise, and best even with a yes or no answer. (2) Personal acquaintance with the problem, having found it not easy. (3) A belief that the question, its solution, partial results or even attempts at its solution are likely to have great importance for the development of the mathematical aspects of Celestial Mechanics.

\end{abstract}

\section*{Introduction}

In February 2005, some participants of the meeting Matemairacorana, organized in Recife, Brazil, by Hildeberto Cabral, met and discussed open questions. In December 2010, an international meeting on Hamiltonian systems and Celestial Mechanics, called Sidimec, took place in Aracaju, Brazil, in honor of Hildeberto Cabral on his 70th birthday. All participants were invited to make contributions to a list of open questions.
The problems deal with the dynamics of equations (\ref{nbp}), invariant manifolds, existence of particular solutions, singularities, relative equilibria and central configurations.

\subsection*{A few words on the $n$-body problem} The $n$-body problem is defined by the second order system of ordinary differential equations 
\begin{equation}\label{nbp}
 \ddot{\mathbf{q}}_i=\sum_{j\ne i}^n m_jr_{ij}^{2a}(\mathbf{q}_j-\mathbf{q}_i)
\end{equation}
where the positions $\mathbf{q}_i$ are in $\mathbf{R}^3$,  the body located at $\mathbf{q}_i$ has mass $m_i$, $r_{ij}=\|\mathbf{q}_i-\mathbf{q}_j\|$ and  $a=-3/2$. In some questions we consider the same equations with different values of $a$. When nothing about $a$ is specified, or when we write ``the Newtonian $n$-body problem", we mean $a=-3/2$. The quantities
\begin{equation}\label{mom}
\mathrm{I}=\frac{1}{M}\displaystyle\sum_{i<j}m_im_jr_{ij}^2,\qquad \mathrm{U}=-\frac{1}{2(a+1)}\displaystyle\sum_{i<j}m_im_jr_{ij}^{2(a+1)},
\end{equation}
where $M$ is the total mass,  $a\neq -1$, are respectively called the moment of inertia and the potential.

\subsection*{A few words on central configurations}  In the $n$-body problem, a configuration is {\it central }if there exists a real number $\lambda$ such that
\begin{equation}\label{cc-eq}
 \lambda (\mathbf{q}_i -\mathbf{q}_G) = \sum_{j\neq i} m_j \|\mathbf{q}_i - \mathbf{q}_j \|^{2a}
(\mathbf{q}_i - \mathbf{q}_j).
\end{equation}
By $\mathbf{q}_G$ we mean the position of the center of mass of
the system. Here the exponent $a$ is often taken less than $-1$. If $a=-3/2$, we are considering the Newtonian force
law of Celestial Mechanics.

Due to the invariance of Newton's equation by rotations, one can define a reduced problem.
A fixed point of the reduced problem is called a \textit{relative equilibrium}. One can prove that the configuration of
a relative equilibrium is a planar central configuration. One counts central configurations or relative equilibria up to
rotation and rescaling.

\section*{The problems}

\paragraph{\bf Problem {1}{} -- Paul Painlev\'e --} Can a non-collision singularity occur in the 4-body problem?

\paragraph{\bf Comments.} This would be a singularity at a time $t_1$ which is not a collision and such that there is no
collision in a neighborhood of $t_1$. The question is implicit in Painlev\'e's {\it Le\c cons de Stockholm}, \S 20 (1895, see Painlev\'e 1897, 1972) ``{\it Si, $t$ tendant vers $t_1$,
certains des $n$ corps ne tendent vers aucune position limite \`a distance finie, il existe au moins quatre corps ${\rm
M}_1,\dots,{\rm M}_\nu$ ($\nu\geq 4$), qui ne tendent vers aucune position limite, et tels que le minimum $\rho(t)$ des
distances mutuelles $r_{ij}$ de ces points ${\rm M}_1,\dots,{\rm M}_\nu$ tende vers z\'ero avec $t-t_1$, sans qu'aucune
des quantit\'es $r_{ij}$ tende constamment vers z\'ero.} La singularit\'e en question ne saurait donc se produire que
par suite de croisements de $\nu$ astres entre eux $(\nu\geq 4)$, croisements de plus en plus fr\'equents quand $t$ tend
vers $t_1$ et de plus en plus semblables \`a des chocs. Ces {\it pseudo-chocs} ont d\'ej\`a \'et\'e signal\'es par M.\ Poincar\'e comme pouvant donner naissance (pour $n>3$) \`a des solutions p\'eriodiques d'une nature particuli\`ere."

Gerver (2003) recalls what is known of the subject of non-collision singularities, in particular the famous results by Mather and McGehee, by Xia and by him. He then suggests a possible solution of the 4-body case. A related question is proposed in Xia (2002).

\paragraph{\bf Problem {2}{} -- Aurel Wintner --} Is symmetry on the masses and the configurations a necessary condition for any flat
but non-planar solution of the $n$-body problem, if $n>3$?

\paragraph{\bf Comments.} This problem is proposed in \S 389bis of Wintner's book. A {\it flat} solution is such that the configuration
is always in a plane. Such solution is called {\it planar} if this plane is fixed. Wintner gives examples of flat solutions which generalize {\it isosceles} solutions, where there is a symmetry. The study of flat solutions is suggested by the following result. If the configuration is always on a line, then the solution is planar. This line is fixed or the configuration is central (Wintner 1941, \S331).

\paragraph{\bf Problem {3}{} --} In the three-body problem with given masses, is there an algebraic invariant subvariety in the phase space besides the known ones?

\paragraph{\bf Comments.} The known ones are obtained by fixing the energy and/or the angular momentum and/or by
restricting to an isosceles problem and/or by restricting to the set of homographic solutions with a given central
configuration and/or by restricting to some lower dimensional 3-body problem. If answered negatively then Bruns'
theorem (see Julliard-Tosel 2000) is reproved, as well as several problems similar to
\textbf{Problem {2}}. In this kind of problems one assumes an algebraic constraint along a solution and tries to prove that it implies a much
stronger algebraic constraint. Taking successive time derivatives of this constraint, one finds a family of constraints that defines an
invariant algebraic manifold. A classification of such manifolds would give another proof of the following results.
\begin{enumerate}
 \item  if the bodies form an isosceles triangle for all time, then either it is equilateral, or two masses are equal and the problem is isosceles (Wintner 1941, \S389).

 \item if a $3$-body solution has a constant moment of inertia, then it is a relative equilibrium (Saari's conjecture, see Moeckel 2008).

 \item  a solution whose plane containing the three bodies makes a constant angle with the plane orthogonal to
the angular momentum vector is either planar or isosceles (Salehani 2011).
\end{enumerate}

\paragraph{\bf Problem {4}{} -- Richard Montgomery --} In the 3-body problem with zero angular momentum and negative energy, are all bi-infinite syzygy sequences realized by collision-free solutions?
Are all periodic sequences realized by periodic solutions? Are all finite syzygy sequences realized by solutions asymptotic to an equilateral triple collision when time increases and when time decreases?

\paragraph{\bf Comments.} In the zero angular momentum, negative energy three-body problem
all solutions but Lagrange's total collapse solution have ``syzygies'': instants where the
three masses are collinear but not coincident (Montgomery 2007). The non-collision syzygies
come in three flavors, 1, 2, 3, depending on which mass is between the other two at the syzygy.
Thus, any collision-free solution to the problem yields a bi-infinite sequence,
such as ..123231122.. in the symbols 1, 2, 3. Simply write the flavors of syzygy
in their temporal order of occurrence. 

If we change the potential from a $1/r$ to a $1/r^2$ potential (Montgomery 2005), syzygy sequences may provide full information on the solutions. If we take the three masses to be equal, the bounded non-collision orbits are in 1:1 correspondence, modulo symmetries, with the bi-infinite sequences where the same syzygy does not appear twice in a row, and where the same pair of syzygies does not appear infinitely many times in a row.

\paragraph{\bf Problem {5}{} -- Alain Chenciner, Andrea Venturelli --} Problem of the infinite spin in the total collision of the $n$-body problem: in the
planar $n$-body problem, may a configuration make an infinite number of revolutions before arriving at a total collapse?

\paragraph{\bf Comments.} Several claims of even stronger results were published, but even in this basic case, we cannot
find a proof in the corresponding papers. Chazy answered negatively if the limiting central configuration is
non-degenerate. Chazy (1918) indeed ``postulates" the non-degeneracy of any central configuration, and uses the postulate in the proof of this result (op.\ cit.\
p.\ 361, footnote 1). This postulate is wrong (see e.g.\ Albouy and Kaloshin 2012). Probably for this reason, Wintner (1941, p.\ 431) rightly considered that Chazy did not discard the infinite spin in the 4-body problem.

\paragraph{\bf Problem {6}{} -- Jean-Pierre Marco --} Is the topological entropy positive, in the dynamics of the isosceles
collisionless 3-body problem, for values of the angular momentum and of the energy such that the integral manifold is a
smooth compact manifold?

\paragraph{\bf Problem {7}{} -- George David Birkhoff, Michel Herman --} Let ${\cal M}$ be an integral manifold of the 3-body problem.
Do solutions for which $\mathrm{I}\to +\infty$ as $t\to +\infty$ fill up ${\cal M}$ densely?

\paragraph{\bf Comments.}  As usual the center of mass is fixed at the origin.  An integral
manifold is the intersection of a level of the energy and a level of the angular momentum. Birkhoff (1927) asks
this question in the difficult and natural case where the angular momentum is non-zero, the energy is negative and the
 dimension is 3. But even the case with a zero angular momentum and thus a planar solution is
unsolved. Note that if the energy is non-negative the answer is yes. Herman (1998) insisted on this
question and reformulated it.

\paragraph{\bf Problem {8}{} -- Giovanni F.\ Gronchi --} Consider the distance $d$ between two points on two distinct confocal
 ellipses in $\mathbf{R}^3$. If both eccentricities are $>0$, is 12 the maximum number of stationary
points of the function $d^2$? If only one of the eccentricities is $0$, can this upper bound be reduced to 10?

\paragraph{\bf Comments.} Kholshevnikov and Vassiliev (1999) conjecture that the answer to the first question is positive. Examples with 12 stationary points, and with 10 points in the circle-ellipse case, were
found (Gronchi 2002). Moreover, the cases with infinitely many points were completely classified (either two
coinciding conics or two concentric coplanar circles) and are excluded here.

\paragraph{\bf Problem {9}{} -- Jean Chazy, Aurel Wintner, Steve Smale --} Is the number of relative equilibria finite, in the
$n$-body problem of Celestial Mechanics, for an arbitrarily given choice of positive real numbers $m_1,\dots,
m_n$ as the masses?
\paragraph{\bf Comments.} See Smale (1998), Hampton and Moeckel (2006), Hampton and Jensen (2011). The last result (Albouy and Kaloshin 2012) on this question is: for 5 bodies in the plane, the number is finite, except perhaps if the masses belong to an explicit codimension 2 subvariety
of the space of positive masses.

\paragraph{\bf Problem {10}{} --} Given four masses $m_1,m_2,m_3, m_4$ and with $a=-3/2$, is there only one convex central
configuration for each cyclic order?
\paragraph{\bf Comments.} MacMillan and Bartky (1932) proved there is at least one. Xia (2004) has a simpler
argument. The uniqueness is known if the two ends of a diagonal carry equal masses (Albouy et al.\ 2008).
It is not known if, the bodies being numbered 1, 2, 3, 4 in cyclic order,  $m_1=m_2$ and $m_3=m_4$.
The last theorem in MacMillan and Bartky (1932, \S18) gives a part of the answer in this case.

\paragraph{\bf Problem {11}{} --} For any $a$ in some interval containing the exponent $-3/2$, is there a unique central
configuration of four equal masses with a given axis of symmetry and no other symmetry?

\paragraph{\bf Comments.} Albouy and Sim\'o (Albouy 1995) conjecture that the answer is positive.

\paragraph{\bf Problem {12}{} --}  For any $a\leq-1$, except for the regular $n$-gon with equal masses, are
there central configurations of $n$ bodies lying on a circle and having their center of mass at the center of the
circle?

\paragraph{\bf Comments.} In the Newtonian case, where $a=-3/2$, and for $n=4$, this problem was answered negatively 
by Hampton (2005). However, we are asking for a general method of proving the symmetry of the central
configurations subjected to constraints on the geometry and the masses, namely, the configuration is co-circular and
the center of mass is at the center of the circle, regardless the value of the exponent $a$. We suggest to take $a=-1$
or $n = 4$ as a good starting point.

\paragraph{\bf Problem {13}{} --} In the Newtonian five-body problem with equal masses, is every central
configuration symmetric?

\paragraph{\bf Comments.} For convex spatial central configurations the answer is yes (Albouy et al.\ 2008). See
Santos (2004) for very related results. According to the computations in Lee and Santoprete (2009),
one can be almost sure of a positive answer.

\paragraph{\bf Problem {14}{} -- Jaume Llibre --}  Consider the planar central configurations of $N$ bodies of mass $\epsilon$ and a body of unit mass.
Consider their non-coalescent limits when $\epsilon\to 0$. If $N \ge 9$, should the
infinitesimal bodies form a regular polygon? If $N \le 8$, are the limiting central configurations necessarily
symmetric?

\paragraph{\bf Comments.}  Non-coalescent means that the infinitesimal bodies all have distinct limiting positions. Hall (1987)
got the first results on this problem, about $N=2$, $N=3$ and very large $N$,  while Salo and Yoder (1988) obtained numerically
a conjecturally complete list of configurations. A positive answer to {\bf Problem {14}} was obtained for all
$N>\exp(73)$ in Casasayas et al.\ (1994), for $N=4$ in Albouy and Fu (2009).

\paragraph{\bf Problem {15}{} -- Rick Moeckel --} In the planar Newtonian $n$-body problem, consider a solution of
relative equilibrium which is linearly stable (i.e.\ a fixed point of the reduced system which is linearly stable). Is
there always a dominant mass, i.e.\ a body with a mass, let us say, at least 10 times bigger than the total mass of the
other bodies?

\paragraph{\bf Comments.} This would imply the instability of any relative equilibrium with equal masses. Even this is not proved, except if $n=3$, 4 or $n\geq 24306$ (Roberts 1999).

\paragraph{\bf Problem {16}{} -- Rick Moeckel --} Under the same hypothesis as in  \textbf{Problem {15}}, is the configuration always a non-degenerate
minimum of the function $\mathrm{U}$ restricted to the sphere $\mathrm{I}=1$?

\paragraph{\bf Comments.} This question is suggested by Theorem 1 in
Moeckel (1994) and by \textbf{Problem {15}}. The central configuration should correspond to a critical point with even index (Hu and Sun 2009).

\paragraph{\bf Problem {17}{} --} Does there exist a planar 5-body central configuration which has a degeneracy in the vertical
direction?

\paragraph{\bf Comments.} We mean a degeneracy of the central configuration seen as a critical point, as in \textbf{Problem {16}}.  Moeckel and Sim\'o (1995) proved that such a degeneracy does occur in the 946-body problem. In the tensorial approach of central configurations due to Albouy and Chenciner (see Albouy 1997, p.\ 72), one defines the corank of a relative central configuration $\beta$ with multiplier $\lambda$ as being the rank of the tensor
$\alpha=dg(\beta)$ where $g$ is the real function
$$g(\beta)=\mathrm{2U}(\beta)+\lambda\mathrm{I}(\beta).$$
One knows that a planar relative central configuration of the five-body problem has corank  $\leq2$.
\textbf{Problem {17}} asks if there is a central configuration of the planar five-body problem with corank one.

\bigskip\bigskip

{\bf Acknowledgments.} The authors thank the referees, Alain Chenciner, Gonzalo Contreras,  Jacques F\'ejoz, Antonio Carlos Fernandes, Giovanni-Federico Gronchi, Marcel Guardia, Jaume Llibre, Jean-Pierre Marco, Rick Moeckel, Richard Montgomery, Gareth Roberts, Alfonso Sorrentino, Andrea Venturelli and Claudio Vidal for having enriched our work with many questions and comments. We benefited from the lists of open questions on the $n$-body problem available on Richard Montgomery's web page. We thank the regional program Math-AmSud for supporting our work.

\section*{References}

\hangindent=2em
\hangafter=1
Albouy, A.: The symmetric central configurations of four equal masses. Contemp.\ Math.\ {\bf 198}, 131--135 (1995)

\hangindent=2em
\hangafter=1
Albouy, A.: Recherches sur le probl\`eme des $n$ corps. Notes scientifiques et techniques du Bureau des Longitudes S058. Institut de M\'ecanique C\'eleste et Calcul des \'Eph\'em\'erides, Paris (1997)

\hangindent=2em
\hangafter=1
Albouy, A., Fu, Y.: Relative equilibria of four identical satellites. Proc.\ R.\ Soc.\ A {\bf 465}(2109), 2633--2645 (2009)

\hangindent=2em
\hangafter=1
Albouy, A., Kaloshin, V.: Finiteness of central configurations of five bodies in the plane. Ann.\ Math.\ {\bf 176}(1), 535--588 (2012)

\hangindent=2em
\hangafter=1
Albouy, A., Fu, Y., Sun, S.: Symmetry of planar four-body convex central configurations. Proc.\ R.\ Soc.\ A {\bf 464}(2093), 1355--1365 (2008)

\hangindent=2em
\hangafter=1
Birkhoff, G.D.: Dynamical Systems, volume IX, p.\ 290. Am.\ Math.\ Soc.\ Colloquium Pub.\ (1927)

\hangindent=2em
\hangafter=1
Casasayas, J., Llibre, J., Nunes, A.: Central configurations of the planar $1 + N$-body problem. Celest.\ Mech.\ Dyn.\ Astron.\ {\bf 60}, 273--288 (1994)

\hangindent=2em
\hangafter=1
Chazy, J.: Sur certaines trajectoires du probl\`eme des $n$ corps. Bull.\ Astron.\ {\bf 35}, 321--389 (1918)

\hangindent=2em
\hangafter=1
Gerver, J.L.: Noncollision singularities: do four bodies suffice? Exp.\ Math.\ {\bf 12}(2), 187--198 (2003)

\hangindent=2em
\hangafter=1
Gronchi, G.F.: On the stationary points of the squared distance between two ellipses with a common
focus. SIAM J.\ Sci.\ Comput.\ {\bf 24}(1), 61--80 (2002)

\hangindent=2em
\hangafter=1
Hall, G.R.: Central Configurations in the Planar $1 + n$ body problem. Pre\-print, Boston University, Boston
(1987)

\hangindent=2em
\hangafter=1
Hampton, M.: Co-circular central configurations in the four-body problem. In: Equadiff 2003 International
Conference on Differential Equations, pp.\ 993--998. World Sci.\ Publ.\ Co.\ Pte.\ Ltd.\ (2005)

\hangindent=2em
\hangafter=1
Hampton, M., Jensen, A.: Finiteness of spatial central configurations in the five-body problem. Celest.\ Mech.\ Dyn.\ Astron.\ {\bf 109}, 321--332 (2011)

\hangindent=2em
\hangafter=1
Hampton, M., Moeckel, R.: Finiteness of relative equilibria of the four-body problem. Invent.\ Math.\ {\bf 163}, 289--312 (2006)

\hangindent=2em
\hangafter=1
Herman, M.: Some open problems in dynamical systems. In: Proceedings of the International Congress of Mathematicians, Documenta Mathematica J.\ DMV, Extra volume ICM II, pp.\ 797--808 (1998)

\hangindent=2em
\hangafter=1
Hu, X., Sun, S.: Stability of relative equilibria and Morse index of central configurations. Comptes Rendus Mathematique {\bf 347}(21--22), 1309--1312 (2009)

\hangindent=2em
\hangafter=1
Julliard-Tosel, E.: Bruns' theorem: The proof and some generalizations. Celest.\ Mech.\ Dyn.\ Astron.\ {\bf 76}, 241--281 (2000)

\hangindent=2em
\hangafter=1
Kholshevnikov, K.V., Vassiliev, N.N.: On the distance function between two Keplerian elliptic orbits. Celest.\ Mech.\ Dyn.\ Astron.\ {\bf 75}, 75--83 (1999)

\hangindent=2em
\hangafter=1
Lee, T.-L., Santoprete, M.: Central configurations of the five-body problem with equal masses. Celest.\ Mech.\ Dyn.\ Astron.\ {\bf 104}(4), 369--381 (2009)

\hangindent=2em
\hangafter=1
MacMillan, W.D., Bartky, W.: Permanent configurations in the problem of four bodies. Trans.\ Am.\ Math.\ Soc.\ {\bf 34}(4), 838--875 (1932)

\hangindent=2em
\hangafter=1
Moeckel, R.: Linear stability of relative equilibria with a dominant mass. J.\ Dyn.\ Differ.\ Equ.\ {\bf 6}(1), 37--51 (1994)

\hangindent=2em
\hangafter=1
Moeckel, R.: A proof of Saari's conjecture for the three-body problem in $\mathbf{R}^d$. Disc.\ Cont.\ Dyn.\ Sys.\ Ser.\ S {\bf 1}(4), 631--646 (2008)

\hangindent=2em
\hangafter=1
Moeckel, R., Sim\'o, C.: Bifurcation of spatial central configurations from planar ones. SIAM J.\ Math.\ Anal.\ {\bf 26}(4), 978--998 (1995)

\hangindent=2em
\hangafter=1
Montgomery, R.: Fitting hyperbolic pants to a three-body problem. Ergod.\ Theory Dyn.\ Syst.\ {\bf 25}(3), 921--947 (2005)

\hangindent=2em
\hangafter=1
Montgomery, R.: The zero angular momentum, three-body problem: all but one solution has syzygies. Ergod.\ Theory Dyn.\ Syst.\ {\bf 27}(6), 1933--1946 (2007)

\hangindent=2em
\hangafter=1
Painlev\'e, P.: Le\c cons sur la th\'eorie analytique des \'equations diff\'erentielles. Hermann, Paris (1897)

\hangindent=2em
\hangafter=1
Painlev\'e, P.: \OE uvres, Tome 1. \'Editions du CNRS, Paris (1972)

\hangindent=2em
\hangafter=1
Roberts, G.E.: Spectral instability of relative equilibria in the planar $n$-body problem. Nonlinearity {\bf 12},
757--769 (1999)

\hangindent=2em
\hangafter=1
Salehani, M.K.: Global geometry of non-planar 3-body motions. Celest.\ Mech.\ Dyn.\ Astron.\ {\bf 111}, 465--479 (2011)

\hangindent=2em
\hangafter=1
Salo, H., Yoder, C.F.: The dynamics of coorbital satellite systems. Astron.\ Astrophys.\ {\bf 205}, 309--327 (1988)

\hangindent=2em
\hangafter=1
Santos, A.A.: Dziobek's configurations in restricted problems and bifurcation. Celest.\ Mech.\ Dyn.\ Astron.\ {\bf 90}(3--4), 213--238 (2004)

\hangindent=2em
\hangafter=1
Smale, S.: Mathematical problems for the next century. Math.\ Intell.\ {\bf 20}(2), 7--15 (1998)

\hangindent=2em
\hangafter=1
Wintner, A.: The Analytical Foundations of Celestial Mechanics. Princeton University Press, Princeton (1941)

\hangindent=2em
\hangafter=1
Xia, Z.: Some of the problems that Saari didn't solve. Contemp.\ Math.\ {\bf 292}, 267--270 (2002)

\hangindent=2em
\hangafter=1
Xia, Z.: Convex central configurations for the $n$-body problem. J.\ Differ.\ Equ.\ {\bf 200}(2), 185--190 (2004)

\end{document}